\documentstyle[12pt]{article}
\begin{document}
\setlength{\baselineskip}{0.3in}
\begin{flushright}
DPNU-94-56\\
INS-rep.-1075,\\
Nov.1994
\end{flushright}

\begin{center}

{\Large {\bf Can the Nambu-Goldstone Boson Live\\
on the Light-Front?}}
\vspace{25pt}
\noindent

Yoonbai Kim\hspace{2pt}\footnote{E-mail address:
yoonbai@eken.phys.nagoya-u.ac.jp},
Sho Tsujimar$\mbox{u}^{*}$\footnote{E-mail
address: sho@ins.u-tokyo.ac.jp}
and
Koichi Yamawaki\hspace{2pt}\footnote{E-mail
address: yamawaki@eken.phys.nagoya-u.ac.jp}\\
\vspace{18pt}
\noindent
Department of Physics,  Nagoya University,
Nagoya 464-01, Japan \\
$\mathop{}^{*}$Institute for Nuclear Study,
University of Tokyo, Tanashi, Tokyo 188, Japan
\end{center}
\vspace{15pt}
\begin{abstract}
We show that the Nambu-Goldstone(NG) boson restricted on the
light-front(LF) can only exist if we regularize the theory by
introducing the explicit
breaking NG-boson mass $m_{\pi}$. The NG-boson zero mode, when
integrated over the LF, must have a singular
behavior $\sim 1/m_{\pi}^2$ in the symmetric limit of 
$m_{\pi}^2\rightarrow 0$.
In the discretized LF quantization this peculiarity is clarified
in terms of the zero-mode constraints in the linear $\sigma$ model.
The LF charge annihilates the vacuum, while it is not conserved
in the symmetric limit in the NG phase.
\end{abstract}
\clearpage
Recently there has been renewed interest in the light-front (LF)
quantization \cite{Dira} as a promising approach to solve the
nonperturbative dynamics \cite{CHA,Wils}. Based on the trivial
vacuum structure, the LF quantization with a Tamm-Dancoff truncation
has successfully described the bound state spectra and their wave
functions in several field theoretical models in (1+1) dimensions,
particularly within the framework of the discretized LF quantization
(DLFQ) \cite{MY,PB}. However, realistic theories like QCD in (3+1)
dimensions include rich structures such as confinement,
spontaneous symmetry breaking (SSB), etc., which are basically
on account of the nontrivial vacuum in the conventional
equal-time quantization.
How can one reconcile such a nontrivial structure of the
theory with the trivial vacuum of the LF quantization?
It seems to be now a
general consensus that the zero mode \cite{MY} plays an essential role
to realize the spontaneous symmetry breaking on the LF \cite{H,Rob,Wils}.
Problem of the zero mode in the LF vacuum was first addressed
back in 1976 by Maskawa and Yamawaki \cite{MY} who discovered,
 within the canonical theory of DLFQ, the second class
 constraint so-called zero mode
constraint, through which the zero mode is not an independent degree of
freedom but a complicated operator-valued function of all other
modes.
One may thus expect that solving the vacuum state in the
ordinary equal-time quantization is traded for solving the
operator zero mode in the LF quantization.
Actually, several authors have recently argued in (1+1)
dimensional models that the zero-mode solution might induce the spontaneous
breaking of discrete symmetries \cite{Rob}. However,
the most outstanding feature of the spontaneous symmetry breaking is
the existence of the Nambu-Goldstone (NG) boson for the continuous symmetry
breaking. Thus the real question to be addressed is whether
or not the zero-mode solution automatically produces
the NG phase, particularly in (3+1) dimensions.

In this paper we shall show, in the context of DLFQ, how  the NG
phenomenon is realized due to the zero modes in (3+1) dimensions
while the vacuum remains in the trivial LF vacuum.
We encounter a striking feature of the
zero mode of the NG boson: Naive use of the zero-mode constraints
does not lead to the NG phase at all (``no-go theorem'') in contrast
to the current expectation mentioned above.
Within the DLFQ, it is inevitable to introduce an infrared regularization by
explicit breaking mass of the NG boson $m_\pi$.
The NG phase can only be realized via peculiar
behavior of the zero mode of the NG-boson fields:
The NG-boson zero mode, when integrated over the LF,
must have a singular behavior $\sim 1/m_{\pi}^2$ in the symmetric limit
$m_\pi^2 \rightarrow 0$.
This we demonstrate both in
a general framework of the LSZ reduction formula
and in a concrete field theoretical model,
the linear $\sigma$ model, within a framework of DLFQ.
The NG phase is in fact realized in such a way that the vacuum is trivial
while the LF charge is {\it not conserved} in the symmetric limit
$m_\pi^2 \rightarrow 0$.

Let us first prove a ``no-go theorem''
that the naive LF restriction of the NG-boson field
leads to vanishing of both the NG-boson emission vertex
and the corresponding current vertex; namely, the NG phase
is not realized in the LF quantization.

Based on the LSZ reduction formula,
the NG-boson emission vertex $A \rightarrow B + \pi$ may be written as
\begin{eqnarray}
\lefteqn{\langle B \pi(q)\vert A \rangle = i\int d^4 x e^{iqx} \langle B
\vert \Box \pi(x) \vert A \rangle} \nonumber \\
&=&i(2\pi)^4 \delta(p_A^{-}-p_B^{-}-q^-)
\delta^{(3)}(\vec{p}_A-\vec{p}_B-\vec{q})
\langle B \vert j_{\pi}(0)\vert A
\rangle ,
\end{eqnarray}
where $\pi(x)$ and $j_{\pi}(x)=\Box \pi(x)=(2\partial_{+}\partial_{-}
-\partial_{\bot}^2) \pi(x)$ are
the interpolating field and the source function of the NG boson,
respectively, and $q^{\mu}=p^{\mu}_A-p^{\mu}_B$
are the NG-boson four momenta and
$\vec{q}\equiv (q^{+},q^{\bot})$ \cite{notation}.
It is customary \cite{Wei} to take the collinear momentum,
 $\vec{q}=0$ and $q^{-}\ne 0$ (not a soft momentum),
for the emission
vertex of the exactly massless
NG boson with $q^2=0$.
Here we adopt the DLFQ, $x^{-}\in [-L,L]$,  with a periodic boundary
condition \cite{lfc}
in the $x^{-}$ direction  and take the continuum limit
$L \rightarrow \infty$
in the end of the whole calculation \cite{MY}.
Without specifying
the boundary condition, we would not be able to
formulate consistently the LF quantization anyway
even in the continuum theory \cite{STH}.
Then the NG-boson emission vertex should vanish on the LF
due to the periodic boundary condition:
\begin{eqnarray}
\lefteqn{(2\pi)^3\delta^{(3)}(\vec{p}_A-\vec{p}_B)
\langle B\vert j_{\pi}(0)\vert
A\rangle}\nonumber\\
&=&\int d^{2}x^{\bot}\lim_{L\rightarrow\infty}\langle B\vert
\Bigl(\int^{L}_{-L}dx^{-}2\partial_{-}\partial_{+}\pi\Bigr)\vert A
\rangle=0.
\end{eqnarray}

Another symptom of this disease is the vanishing of the
current vertex (analogue of $g_A$ in the nucleon matrix element).
When the continuous symmetry is
spontaneously broken, the NG theorem requires that the corresponding
current $J_{\mu}$ contains
an interpolating field of the NG boson $\pi(x)$, that is,
 $J_{\mu}=-f_{\pi}\partial_{\mu}\pi+\widehat
J_{\mu}$, where $f_{\pi}$ is the ``decay constant'' of
the NG boson
and $\widehat J_{\mu}$ denotes the non-pole term.
Then the current conservation $\partial_{\mu} J^{\mu}=0$
leads to
\begin{eqnarray}
\lefteqn{0=
\langle B \vert
\int d^3\vec{x}\, \partial_{\mu}\widehat J^{\mu}(x)\vert A
\rangle_{x^+=0}}\nonumber\\
&=&-i(2\pi)^3\delta^{(3)}(\vec{q})\, \displaystyle{
\frac{m_{A}^2-m_{B}^2}{2p_A^+}}\langle B \vert \widehat J^+(0)
\vert A \rangle,
\label{c-vertex}
\end{eqnarray}
where $\int d^3\vec{x} \equiv \lim_{L\rightarrow \infty}
\int_{-L}^{L} dx^-d^2 x^{\bot}$ and the integral of
the NG-boson sector $\Box\pi$ has no contribution on the LF
because of the periodic boundary condition as we mentioned before.
Thus the current vertex $\langle B\vert\widehat J^{+}(0)\vert A\rangle$
should vanish at $q^2=0$ as far as $m^2_{A}\ne m^2_{B}$
\cite{nucleon}.

This is actually a manifestation of the conservation of a charge
$\widehat Q\equiv
\int d^3\vec{x}\, \widehat J^{+}$
which is constructed only from the non-pole term.
Note that $\widehat Q$ is equivalent to
the full LF charge $Q\equiv
\int d^3 \vec{x}\,  J^{+}$, since
the pole part always drops out of
$Q$ due to the integration on the LF, i.e., $Q=\widehat Q$.
Therefore the conservation of $\widehat{Q} $ inevitably
follows from the conservation of $Q$:
$[\widehat Q, P^-]=[Q, P^-]=0$, which in fact implies vanishing
current vertex mentioned above.
This is in sharp contrast to the
charge integrated over usual space
$\mbox{\boldmath$x$}=(x^1,x^2,x^3)$
in the equal-time quantization: $Q^{\rm et} =
\int d^3\mbox{\boldmath$x$} J^0$ is conserved while
$\widehat Q^{\rm et} = \int d^3\mbox{\boldmath$x$} \widehat J^0$ is not.

Thus the NG bosons are totally decoupled, i.e., the NG phase is
not realized on the LF.
Note that this is a direct consequence of the periodic boundary condition
and
the first-order form of
$\Box=2\partial_{+}\partial_{-}-\partial_{\bot}^2$
in $\partial_{\pm}$ in contrast to
the second order form in $\partial_{0}$ in the equal-time quantization.

Now,  we propose to regularize
the theory  by introducing
explicit breaking mass of the NG boson $m_\pi$.
The essence of the NG phase with a small
explicit symmetry breaking can well be described by the old notion of
the PCAC hypothesis:
 $\partial_{\mu}J^{\mu}(x)=f_{\pi}m_{\pi}^2\pi(x)$, with $\pi(x)$
being the interpolating field of the (pseudo-) NG boson $\pi$.
  From the PCAC relation the current
divergence of the non-pole term $\widehat J^{\mu}(x)$
reads $\partial_{\mu}\widehat J^{\mu}(x)
=f_{\pi}(\Box+m_{\pi}^2)\pi(x)=
f_{\pi}j_{\pi}(x).$
Then we obtain
\begin{eqnarray}
\label{PCAC}
\langle B \vert \int d^3\vec{x}\,
\partial_{\mu}\widehat J^{\mu}(x)
\vert A \rangle
&=&f_{\pi}m_{\pi}^2
\langle B \vert 
\int d^3 \vec{x}\, \pi(x)\vert A \rangle  \nonumber \\
&=&
\langle B \vert
\int d^3 \vec{x}\, f_{\pi} j_{\pi}(x)
\vert A \rangle,
\end{eqnarray}
where the integration of the pole term $\Box \pi(x)$ is dropped 
out as before. The second expression of (\ref{PCAC})
 is nothing but the matrix element of 
the LF integration of the $\pi$ zero mode (with $P^+=0$)
$\omega_{\pi} \equiv \frac{1}{2L}\int_{-L}^{L} dx^- \pi(x)$.
Suppose that $\int d^3\vec{x}\, \omega_{\pi} (x)
=\int d^3\vec{x}\, \pi (x)$ is regular when
$m_{\pi}^2\rightarrow 0$. Then this leads to the ``no-go theorem''
 again. Thus in order to have the non-zero NG-boson emission 
 vertex (R.H.S. of (\ref{PCAC}))
  as well as the non-zero current vertex (L.H.S.) at $q^2=0$, 
 the $\pi$ zero mode
$\omega_{\pi}(x)$ must behave as 
\begin{equation}
\int d^3 \vec{x}\, \omega_{\pi}\sim \frac 1{m_{\pi}^2}
\quad (m_{\pi}^2 \rightarrow 0).
\label{omega}
\end{equation}

This situation may be clarified when the PCAC relation is
written in the momentum space:
\begin{equation}
\displaystyle{
\frac{m_{\pi}^2f_{\pi}j_{\pi}(q^2)}{m_{\pi}^2-q^2}=
\partial^{\mu} J_{\mu}(q)=
\frac{q^2f_{\pi}j_{\pi}(q^2)}{m_{\pi}^2-q^2}
+\partial^{\mu}\widehat J_{\mu}(q).
}
\label{PCAC-mom}
\end{equation}
What we have done when we reached the ``no-go theorem'' can be
summarized as follows.
We first set L.H.S of (\ref{PCAC-mom}) to
zero (or equivalently, assumed implicitly the regular
behavior of $\int d^3\vec{x}\, \omega_{\pi}(x)$)
in the symmetric limit
in accord with the current conservation
$\partial^{\mu} J_{\mu}=0$. Then
in the LF formalism with $\vec{q}=0 $ $(q^2=0)$,
the first term (NG-boson pole term) of R.H.S. was also zero
due to the periodic boundary
condition or the zero-mode constraint.
Thus we arrived at  $\partial^{\mu}\widehat J_{\mu}(q)=0$.
However, this procedure is
equivalent to playing a 
nonsense game: $\lim_{m_{\pi}^2,\,q^{2}\rightarrow 0}
(\frac{m^{2}_{\pi}-q^{2}}{
m^{2}_{\pi}-q^{2}})=0$ as far as $f_{\pi}j_{\pi}\ne 0$ (NG phase).
Therefore the ``$m_{\pi}^2 =0$'' theory with vanishing L.H.S. 
is ill-defined on the LF, namely,
the ``no-go theorem'' is false.
The correct procedure should be to 
take the symmetric limit $m_{\pi}^2 \rightarrow 0$
{\it after} the LF restriction $\vec{q}=0$ $ (q^2=0)$
\cite{CHKT},
although (\ref{PCAC-mom}) itself yields the same result
$f_{\pi} j_{\pi} = \partial^{\mu} \widehat J_{\mu}$,
irrespectively of the order of the two limits
$q^{2}\rightarrow 0$ and
$m_{\pi}^2\rightarrow 0$. Then  (\ref{omega}) does follow.
This implies that at quantum level the LF charge $Q=\widehat Q$ 
 is not conserved, or the current conservation does not hold
for a particular Fourier component with $\vec{q}=0$ even 
in the symmetric limit: 
\begin{equation}
\frac{1}{i}[Q, P^{-}]=\partial^{\mu} J_{\mu}\vert_{\vec{q}=0}=f_{\pi} 
\lim_{m_{\pi}^2\rightarrow 0}
m^2_{\pi}\int d^3 \vec{x}\, \omega_{\pi} \neq 0.
\label{nonconserv}
\end{equation}

Let us now demonstrate that   
 (\ref{omega}) and (\ref{nonconserv}) indeed take place {\it
 as the solution
 of the constrained zero-modes} in the NG phase
 of the $O(2)$ linear $\sigma$ model:  
\begin{equation}\label{lag}
{\cal L}=\frac{1}{2}(\partial_{\mu}\sigma)^2+\frac{1}{2}
(\partial_{\mu}\pi)^2-\frac{1}{2}\mu^2 (\sigma^2+\pi^2)-
\frac{\lambda}{4}
(\sigma^2+\pi^2)^2 +c\sigma,
\end{equation}
where the last term is the explicit breaking
which regularizes the NG-boson zero mode.

In the DLFQ we can clearly separate
the zero modes (with $P^{+}=0$), $\pi_0
\equiv \frac{1}{2L}\int_{-L}^{L} dx^- \pi(x)$
(similarly for $\sigma_0$),
from other oscillating modes (with $P^{+} \ne 0$),
$\varphi_{\pi}\equiv \pi-\pi_0$ (similarly
for $\varphi_{\sigma}$).
Through the Dirac quantization of the constrained system
the canonical commutation relation for the 
oscillating modes reads \cite{MY}
\begin{equation}
\left[\varphi_i(x),\varphi_j(y)\right]
=-{i \over 4}\left\{\epsilon(x^--y^-)-{x^--y^-
\over L}\right\}\delta_{ij}\delta^{(2)}(x^{\bot}-y^{\bot}), 
\label{commutator}
\end{equation}
where each index stands for $\pi$ or $\sigma$, and the $\epsilon(x)$
is the sign function. 
By use of (\ref{commutator}) we can introduce 
creation and annihilation operators simply defined 
by the Fourier expansion of
$\varphi_{i}$ with respect to $x^{-}$ even when the interaction is 
included. 
Thus the physical Fock space 
is constructed upon the LF vacuum (``trivial vacuum'') 
which is defined to be annihilated by the annihilation operators
without recourse to the dynamics.

On the other hand, the zero modes 
 are  not independent degrees of freedom but are
implicitly determined by $\varphi_{\sigma}$ and $\varphi_{\pi}$
through the second class constraints so-called 
zero-mode constraints \cite{MY}:
\begin{equation}
\chi_{\pi}\equiv
\displaystyle{\frac 1{2L} \int^L_{-L}dx^-}
\left[(\mu^2-\partial_{\bot}^2)\pi
+\lambda \pi(\pi^2+\sigma^2)\right]= 0,
\label{pizero}
\end{equation}
and similarly,  $\chi_{\sigma}\equiv 
\frac{1}{2L}\int^L_{-L}dx^-\{[\pi \leftrightarrow \sigma]-c\} = 0$.
Note that through the equation of motion  
these constraints are equivalent  to
the characteristic of the DLFQ with periodic boundary condition:
$
\chi_{\pi}=-\frac{1}{2L}\int^L_{-L}dx^-\,2\partial_{+}\partial_{-}\pi
=0,
$ (similarly for $\sigma$)
which we have used to prove the ``no-go theorem'' for the case of
$m_{\pi}^2\equiv 0$.

Actually, in the NG phase $(\mu^2 < 0)$  the equation of motion 
of $\pi$ reads
$(\Box+m_{\pi}^2)\pi(x)=-\lambda({\pi}^3+\pi
{\sigma}'^2+2v\pi{\sigma}') $ $\equiv j_{\pi}(x)
$, with ${\sigma}'=
\sigma-v$ and $m_{\pi}^2=\mu^2+\lambda v^2=c/v$, 
where $v\equiv \langle \sigma\rangle$ is the classical 
vacuum solution  determined by  $\mu^2 v
+\lambda v^3 =c$. 
Integrating the above equation of motion over $\vec{x}$, we have
\begin{equation}
\int d^{3} \vec{x}\, j_{\pi}(x) -m_{\pi}^2 \int d^{3} \vec{x}\, 
\omega_{\pi}(x) 
=\int d^{3} \vec{x}\, \Box\pi(x)
=-\int d^{3} \vec{x}\, \chi_{\pi}=0,
\label{eqmot}
\end{equation} 
where $\int d^{3} \vec{x}\, \omega_{\pi}(x) =
\int d^{3} \vec{x}\, \pi(x)$.
Were it not for the singular behavior
(\ref{omega}) for the $\pi$ zero mode
$\omega_{\pi}$, we would have concluded 
$ (2\pi)^3\delta^{(3)}(\vec{q})\,
\langle \pi \vert j_{\pi}(0) \vert \sigma \rangle=
-\langle \pi \vert \int d^3 \vec{x}\, \chi_{\pi}
\vert \sigma \rangle=0$
in the symmetric limit  $m_{\pi}^2 \rightarrow 0$.
Namely, the NG-boson vertex at $q^2=0$ would have vanished, which is
exactly what we called ``no-go theorem''
now related to the zero-mode constraint $\chi_{\pi}$.
On the contrary, direct evaluation of the matrix element of
$j_{\pi}=-\lambda({\pi}^3+\pi
{\sigma}'^2+2v\pi{\sigma}') $ 
in the lowest order perturbation yields non-zero result 
even in the symmetric limit $m_{\pi}^2\rightarrow 0$:
$\langle \pi \vert j_{\pi}(0) \vert \sigma \rangle
=-2\lambda v \langle \pi \vert \varphi_{\sigma} \varphi_{\pi}\vert
\sigma\rangle =-2 \lambda v \ne 0 \quad(\vec{q}=0),
$

which is in agreement with the usual equal-time formulation.
Thus we have seen that  
naive use of the zero-mode constraints by setting $m_{\pi}^2\equiv 0$ 
leads to the internal inconsistency in the NG phase.
The ``no-go theorem'' is again false.

We now study the 
solution of the zero-mode
constraints in the perturbation 
around the classical (tree level) SSB vacuum, since
we need to formulate the NG phase on the LF at least for the theory
whose SSB is already described at the tree level 
in the equal-time quantization.
It is convenient to divide the zero modes
$\pi_{0}$ (or $\sigma_{0}$) into classical constant piece $v_{\pi}$
(or $v_{\sigma}$) and operator part $\omega_{\pi}$ (or
$\omega_{\sigma}$), and also do the zero-mode constraints.
The classical part of the  zero-mode constraints is nothing but the
condition that determines the minimum of the classical potential and
we have chosen a solution that $v_{\pi}=0$ and $v_{\sigma}\equiv v$;
i.e., $\pi_{0}=\omega_{\pi}$, $\sigma_{0}=\omega_{\sigma}+v$.
The operator zero modes
are solved perturbatively by substituting the expansion
$\omega_i=\sum_{k=1}\lambda^k \omega_i^{(k)}$ under the Weyl ordering.

 The lowest order solution of the zero-mode constraints $\chi_{\pi}$
 and $\chi_{\sigma}$ for $\omega_{\pi}$ takes the form:
\begin{equation}
(-m_{\pi}^2+\partial_{\bot}^2)\, \omega_{\pi}
=\frac{\lambda}{2L}\int_{-L}^{L}dx^-(\varphi_{\pi}^3
+\varphi_{\pi}\varphi_{\sigma}^2+2v\varphi_{\pi}\varphi_{\sigma}).
\label{operatorzero}
\end{equation}
Then (\ref{omega}) immediately follows \cite{limit}:
\begin{equation}
\lim_{m_{\pi}^2\rightarrow 0} m_{\pi}^2\int d^3 \vec{x}\, \omega_{\pi}
=-\lambda\int d^3 \vec{x}\, (\varphi_{\pi}^3
+\varphi_{\pi}\varphi_{\sigma}^2+2v\varphi_{\pi}\varphi_{\sigma})
\ne 0.
\label{omega2}
\end{equation}
This is our main result.  This actually ensures non-zero 
 $\sigma \rightarrow \pi \pi$ vertex through (\ref{eqmot}): 
$ 
\langle \pi \vert j_{\pi}(0) \vert \sigma \rangle
=-2\lambda v, 
$
which agrees with the previous direct evaluation as it should.

Let us next discuss the LF charge operator.
The $O(2)$ current in this model is given by $J_{\mu}=\partial_{\mu}
\sigma \pi-\partial_{\mu}\pi \sigma$. As was noted in Ref.\cite{MY}, 
the corresponding LF charge
$
Q=\widehat Q=
\int d^3\vec{x}\,
(\partial_{-}\varphi_{\sigma}\varphi_{\pi}-\partial_{-}
\varphi_{\pi}\varphi_{\sigma})
$
{\it contains no zero-modes} including the $\pi$ pole term which was
dropped by the integration due to
the periodic boundary condition and the $\partial_{-}$,
so that $Q$ is well-defined even in the NG phase and hence 
annihilates the vacuum simply by the $P^+$ conservation:
\begin{equation}
Q \vert  0 \rangle=0.
\label{vac-annih}
\end{equation}    
This is also consistent with explicit computation of the
commutators: 
$\langle [Q, \varphi_{\sigma}]\rangle =-i \langle\varphi_{\pi}\rangle=0$
and 
$\langle [Q,\varphi_{\pi}]\rangle =i\langle 
\varphi_{\sigma}\rangle=0$ \cite{vacc},
 which are contrasted to those in the usual equal-time case 
 where the spontaneously broken charge does not 
annihilate the vacuum: $\langle [Q^{\rm et},\sigma]\rangle
=-i\langle \pi\rangle=0,
\langle [Q^{\rm et},\pi]\rangle=i\langle \sigma\rangle\ne 0$.

Since the PCAC relation is now an operator relation for
the canonical field $\pi(x)$ with $f_{\pi}=v$ in this model, 
(\ref{omega2}) 
 ensures $[\widehat Q,P^-]\ne 0$ or a non-zero 
 current vertex $\langle\pi\vert \widehat J^{+} \vert
\sigma\rangle \ne 0$ 
$ (q^2=0)$ in the symmetric limit.
 Noting that $Q=\widehat Q $, we conclude that  
the regularized zero-mode
constraints indeed lead to non-conservation of the LF charge in the
symmetric limit $m_{\pi}^2\rightarrow 0$:
\begin{equation}
[Q, P^-]=iv 
\lim_{m_{\pi}^2\rightarrow 0} m_{\pi}^2
\int d^3\vec{x}\, \omega_{\pi}\neq 0.
\end{equation}
This can also be confirmed by direct computation of $[Q, P^-]$
through the 
canonical commutator and explicit 
use of the regularized zero-mode constraints.
 At first sight there seems to be 
no distinction between the spontaneous and the explicit symmetry
breakings on the LF. However, the singular behavior of the
NG-boson zero mode 
(\ref{omega}) or (\ref{omega2}) may be understood as a 
characterization of the spontaneous symmetry breaking.

Our result implies that solving the zero-mode constraints 
{\it without regularization}
 would {\it not} lead to the NG phase at all in contradiction
to the naive
 expectation \cite{Rob}.
Our treatment of the zero modes in the canonical DLFQ is quite
different from that proposed recently by Wilson et al. \cite{Wils}
who eliminate the zero modes by hand in the continuum
theory instead
of solving 
the zero-mode constraints.
They also arrived at the non-conservation of the LF charge
{\it without 
zero mode},
while still claiming the conservation of the full LF charge 
in contrast with our result.
 The relationship between these two approaches are not clear at the
moment. Finally, it should be noted that
there exists another no-go
theorem  that forbids any LF field theory (even the
free theory) satisfying the Wightman
axioms \cite{NY}.  This no-go theorem is
also related to the zero modes but 
has not yet been overcome by the DLFQ or any
other existing approach
and is beyond the scope of this paper.

We would like to thank T. Kugo, Y. Ohnuki and I. Tsutsui
for useful discussions.
Y.K. is a JSPS Postdoctoral Fellow (No.93033).
 This work was supported in part by a
Grant-in-Aid for Scientific Research from the
Ministry of Education, Science and Culture (No.05640339), the
Ishida Foundation and the Sumitomo Foundation.
Part of this work was done while K.Y. was staying at
Institute for Theoretical
 Physics at U.C. Santa Barbara in May, 1994,
which was supported in part 
 by the Yoshida Foundation for Science and Technology
 and by the U.S. National Science Foundation under Grant No.PHY89-04035.



\begin{thebibliography}{99}
\bibitem{notation}
We choose the LF ``time'' as $x^{+}=x_{-}\equiv \frac{1}{\sqrt{2}}(x^{0}
+x^{3})$ and the longitudinal and transverse coordinates are
denoted by $\vec{x} \equiv (x^{-},x^{\bot})$ with $x^{-}\equiv
\frac{1}{\sqrt{2}}(x^{0}-x^{3})$ and $x^{\bot}\equiv
(x^{1},x^{2})$, respectively.
\bibitem{lfc} 
A consistent DLFQ can also be constructed with anti-periodic boundary 
condition, which, however, requires each zero mode to be 
identically zero and hence no
vacuum expectation value of the field, namely, no spontaneous
symmetry breaking.
\bibitem{nucleon}
For the nucleon matrix element with $m_A=m_B=m_N$, (\ref{c-vertex}) 
should read $0= \bar u(p_B) 
\gamma^+ \gamma_5 q^- u(p_A) \cdot g_A(q^2=0)$, with $ \bar u 
\gamma^+\gamma_5 q^- u =2m_N\, \bar u i\gamma_5u \ne 0$ ($\vec{q}=0,
q^-\ne 0$), which implies $g_A (0)=0$.
\bibitem{limit}
As is seen from the canonical commutator,
$\pi(x)$ scales like a dimensionless field under the scale transformation
in the $x^{-}$ direction and hence the zero mode $\omega_{\pi}$ 
is independent of $L$. Thus there is no subtlety between the two  
limits, $L\rightarrow \infty$ and $m_{\pi}^2 \rightarrow 0$.
\bibitem{vacc}
By explicit calculation with a careful treatment of the zero-modes
contribution we can also show that
$\langle [Q, \sigma]\rangle =
\langle [Q,\pi]\rangle =0$.
\end{thebibliography}
\end{document}